\documentclass[twocolumn]{aastex631}

\usepackage{amsmath}
\usepackage{cases}
\usepackage{amsfonts}
\usepackage{amssymb}
\usepackage{bm}
\usepackage{color}
\usepackage{dcolumn}
\usepackage{epsfig}
\usepackage{eucal}
\usepackage{float}
\usepackage{graphicx}
\usepackage{microtype}
\usepackage{subfigure}
\usepackage{tikz}
\usepackage{ulem}
\usepackage{verbatim}
\usepackage{xcolor}
\usepackage{hyperref}
\usepackage{wasysym}
\usepackage{mathrsfs}

\begin{document}

\title{Gaussian processes and effective field theory of $f(T)$ gravity\\
under the $H_0$ tension}

\author{Xin Ren}
\affiliation{Department of Astronomy, School of Physical Sciences, University of 
Science and Technology of China, Hefei, Anhui 230026, China}
\affiliation{CAS Key Laboratory for Researches in Galaxies and Cosmology, School 
of Astronomy and Space Science, University of Science and Technology of China, 
Hefei, Anhui 230026, China}

\author{Sheng-Feng Yan}
\affiliation{Istituto Nazionale di Fisica Nucleare (INFN), Sezione di Milano, 
Via Celoria 16, 20146, Milano, Italy}
\affiliation{DiSAT, Universit$\text{\`a}$ degli Studi dell’Insubria, Via 
Valleggio 11, 22100, Como, Italy}
\affiliation{Department of Astronomy, School of Physical Sciences, University of 
Science and Technology of China, Hefei, Anhui 230026, China}

\author{Yaqi Zhao}
\affiliation{Department of Astronomy, School of Physical Sciences, University of 
Science and Technology of China, Hefei, Anhui 230026, China}
\affiliation{CAS Key Laboratory for Researches in Galaxies and Cosmology, School 
of Astronomy and Space Science, University of Science and Technology of China, 
Hefei, Anhui 230026, China}

\author{Yi-Fu Cai} 
\affiliation{Department of Astronomy, School of Physical Sciences, University of 
Science and Technology of China, Hefei, Anhui 230026, China}
\affiliation{CAS Key Laboratory for Researches in Galaxies and Cosmology, School 
of Astronomy and Space Science, University of Science and Technology of China, 
Hefei, Anhui 230026, China}

\author{Emmanuel N. Saridakis}
\affiliation{National Observatory of Athens, Lofos Nymfon, 11852 Athens, 
Greece}
\affiliation{Department of Astronomy, School of Physical Sciences, University of 
Science and Technology of China, Hefei, Anhui 230026, China}
\affiliation{CAS Key Laboratory for Researches in Galaxies and Cosmology, School 
of Astronomy and Space Science, University of Science and Technology of China, 
Hefei, Anhui 230026, China}

\email{yifucai@ustc.edu.cn}
\email{msaridak@phys.uoa.gr}

\begin{abstract}

  We consider the effective field theory formulation of torsional gravity in a cosmological framework to alter the background evolution. Then we use the latest $H_0$ measurement from the SH0ES Team as well as observational Hubble data from cosmic chronometer (CC) and radial baryon acoustic oscillations (BAO) and we reconstruct the $f(T)$ form in a model-independent way by applying Gaussian processes. Since the special square-root term does not affect the evolution at the background level, we finally summarize a family of functions that can produce the background evolution required by the data. Lastly, performing a fitting using polynomial functions, and implementing the Bayesian Information Criterion (BIC), we find an analytic expression that may describe the cosmological evolution in great agreement with observations. 

\end{abstract}


\section{Introduction}
\label{sec:introduction}

The Hubble constant is a very important physical quantity describing the 
characteristics of the universe expansion. The measurement of its value, as 
well as    the explanation of the nature of the accelerating 
expansion \citep{Riess:1998cb} are issues of  high importance in modern 
cosmology. With the development of detection 
technology, its measurement methods are improving  and the corresponding 
accuracy is  increasing \citep{Riess:2018uxu, 
Reid:2019tiq, Yuan:2019npk}. 
However, there is a problem of inconsistency between the results obtained by 
different measurement methods, which is the famous Hubble tension 
\citep{Planck:2018vyg, DiValentino:2020zio,Shah:2021onj}. The SH0ES Team has 
published the latest direct distance ladder measurements of $H_0$, with the 
baseline determination value of $H_0= 73.04 \pm 1.04 \ \mathrm{km\ s^{-1} 
Mpc^{-1}}$\citep{Riess:2021jrx}. Although there are slight differences using 
different observational samples in local measurements, the difference is close 
to or up to 5$\sigma$ compared with Planck 2018 cosmic microwave background 
(CMB) calculation $H_0= 67.40 \pm 0.50\ \mathrm{km\ s^{-1} Mpc^{-1}}$ based on 
$\Lambda$CDM paradigm \citep{Wong:2019kwg}. This  tension  suggests possible 
deviation from the standard $\Lambda\text{CDM}$ cosmological model, inspiring 
people to explore the physical reasons behind this phenomenon \citep{Abdalla:2022yfr}. In particular, 
since 
the new data show greater tension, this issue has again aroused a lot of 
attention and research 
\citep{DiValentino:2015ola,DiValentino:2017oaw,Yang:2018euj,
Yang:2018qmz,Pan:2019jqh,Elizalde:2020mfs,Benevento:2020fev, Sen:2021wld, 
Ballardini:2021eox,Theodoropoulos:2021hkk, Alestas:2022xxm, Dainotti:2022bzg, 
Lee:2022cyh, Cai:2022dkh,Vagnozzi:2019ezj,Odintsov:2022eqm,Dainotti:2021pqg}.

To address the Hubble tension, various modifications  on the early- and 
late-time cosmology have been applied, including early dark energy models 
\citep{Karwal:2016vyq,Vagnozzi:2021gjh}, extra relativistic species \citep{Gelmini:2019deq}, 
modified late-time dark energy models \citep{Zhao:2017cud}, etc (for a review see
\citep{DiValentino:2021izs}). Besides, as a possible interpretation, modified 
gravitational theories are getting increasing attention 
\citep{CANTATA:2021ktz,Nojiri:2010wj,Desmond:2019ygn,Addazi:2021xuf}. Based on 
General Relativity, we may construct   curvature-based extended gravitational 
theories, with $f(R)$ gravity as an example \citep{DeFelice:2010aj}. 
Alternatively, if we start from Teleparallel Equivalent of General Relativity 
(TEGR) \citep{DeAndrade:2000sf, Unzicker:2005in, Aldrovandi:2013wha, 
Krssak:2018ywd}, we will obtain a family of torsion-based modified gravity 
\citep{Cai:2015emx, Krssak:2015oua}. These torsional gravities provide new 
possible mechanisms for cosmological observations, such as inflation and 
accelerated expansion, and are highly considered and widely studied 
\citep{Chen:2010va,Cai:2011tc, Capozziello:2011hj, Bahamonde:2015zma, 
Hohmann:2017jao,Golovnev:2018wbh,Bahamonde:2019zea,Bahamonde:2020cfv, Xu:2018npu,
Jimenez:2020ofm, Santos:2021ypw, Ren:2021uqb,Li:2021mdp, Li:2022mti, Duchaniya:2022rqu, 
Santos:2021ypw,Bahamonde:2021gfp,Bahamonde:2022ohm,Zhang:2021kqn}. It is worth noting that 
these modifications 
are efficient under confrontation with 
galaxy-scale observations too \citep{Chen:2019ftv, Pfeifer:2021njm}.

In the investigation of various modified gravity theories,  Gaussian processes 
reconstruction and effective field theory (EFT) enable us to make a data-driven 
and model-independent analysis. By 
parameterizing a theory with a series of effective parameters or functions, 
effective field theory allows for a systematic investigation of the background 
and perturbations separately, working as a bridge between specific theoretical 
models and   observations. The concept of EFT has been widely applied 
to cosmological studies \citep{Cheung:2007st, Arkani-Hamed:2007ryv, 
Gubitosi:2012hu, Bloomfield:2012ff, Gleyzes:2013ooa,Gong:2022tfu,Frusciante:2019xia,Mylova:2021eld,Gong:2022tfu}, and this approach was 
developed recently for torsional gravity \citep{Li:2018ixg, Cai:2018rzd}. On the 
other hand, the Gaussian processes regression provides us a reliable way to 
obtain fitting functions directly from   observational data, and   it has 
been widely used to reconstruct non-linear functions \citep{Seikel:2013fda, 
Cai:2015zoa, Wang:2017jdm, Elizalde:2018dvw, Aljaf:2020eqh, Holsclaw:2010sk, 
Benisty:2020kdt, Bernardo:2021qhu, Jesus:2021bxq, Rodrigues:2021wyk, 
LeviSaid:2021yat, Cai:2015pia, Mukherjee:2020vkx, vonMarttens:2020apn,Benisty:2022psx}. With 
this approach, we are able to analyse Hubble parameter observational data 
without any special assumption or specific model.

In this work, we analyse  and discuss the Hubble parameter observations from 
the 
perspective of torsional gravity, making use of Gaussian process regression and 
effective field theory. Moreover, we provide one concrete model reconstruction 
in $f(T)$ cosmology. The outline of this work is as follows. In Section 
\ref{sec:EFT review}  we briefly review the EFT of torsional gravity and its 
application in $f(T)$ cosmology. In Section \ref{sec:reconstruction}  we 
reconstruct the evolutionary history of Hubble function with observational 
Hubble data, and we reconstruct the function $f$ with respect to $T$. In 
Section \ref{sec:features} we perform a comparison between the background 
evolution of 
the reconstructed model and standard cosmology. Finally, Section 
\ref{sec:conclusion} is devoted to discussion and conclusions.

\section{The effective field theory approach}
\label{sec:EFT review}

In this section we first briefly review the general effective field theory 
approach, and   we focus on torsional gravity. Then we apply it in the 
framework of $f(T)$ cosmology.

\subsection{EFT of Torsional Gravity}

We start with a brief introduction to the background evolution of the universe 
  from the EFT viewpoint  \citep{Cheung:2007st, Arkani-Hamed:2007ryv, 
Gubitosi:2012hu, Bloomfield:2012ff, Gleyzes:2013ooa}. The EFT action in the 
FLRW metric $d s^{2}=d t^{2}-a^{2}(t) \delta_{i j} d x^{i} d x^{j}$, with $a(t)$ 
the scale factor, for a 
general curvature-based gravity is given by  \citep{Arkani-Hamed:2007ryv}
\begin{align}
\label{eq:genEFTact}
S &= \int d^4x \Bigg\{ \sqrt{-g} \bigg[ \frac{M^2_P}{2} \Psi(t)R - \Lambda(t) - 
b(t)g^{00}
\nonumber
\\
& +M_2^4(\delta g^{00})^2 -\bar{m}^3_1 \delta g^{00} \delta K -\bar{M}^2_2 
\delta K^2 -\bar{M}^2_3
\delta K^{\nu}_{\mu} \delta K^{\mu}_{\nu} \nonumber \\
& +\! m^2_2 h^{\mu\nu}\partial_{\mu} g^{00}\partial_{\nu}g^{00} 
\!+\!\lambda_1\delta R^2\!
+\!\lambda_2\delta R_{\mu\nu}\delta R^{\mu\nu} \!+\!\mu^2_1 \delta g^{00} 
\delta R \bigg]
\nonumber \\
& +\gamma_1 C^{\mu\nu\rho\sigma} C_{\mu\nu\rho\sigma} +\gamma_2
\epsilon^{\mu\nu\rho\sigma} C_{\mu\nu}^{\quad\kappa\lambda} 
C_{\rho\sigma\kappa\lambda}
\nonumber \\
& +\sqrt{-g} \left[ \frac{M^4_3}{3}(\delta g^{00})^3 -\bar{m}^3_2(\delta 
g^{00})^2 \delta K + ... \right] \Bigg\} ~,
\end{align}
where $M_P = (8\pi G_N)^{-{1}/{2}}$ is the reduced Planck mass and $G_N$ the 
Newtonian constant. $R$ is the Ricci scalar corresponding to the 
Levi-Civit$\grave{\mathrm{a}}$ connection, $C^{\mu\nu\rho\sigma}$ is the Weyl 
tensor, $\delta K^{\nu}_{\mu}$ is the perturbation of the extrinsic curvature, 
and $\Psi(t)$, $\Lambda(t)$, $b(t)$ are functions of time depending on the 
background evolution of the universe.

One can apply the EFT approach in the presence of torsion, and extend 
 Eq.~\eqref{eq:genEFTact}  as \citep{Li:2018ixg}
\begin{align}
\label{eq:torEFTact}
S = & \int d^4x \sqrt{-g} \Big[ \frac{M^2_P}{2} \Psi(t)R\! -\! \Lambda(t)\! -\! 
b(t) g^{00}\! +\!
\frac{M^2_P}{2} d(t) T^0 \Big] \nonumber \\
& + S^{(2)} ~.
\end{align}
Comparing to the effective action of curvature-based gravity shown in
 Eq.~\eqref{eq:genEFTact}, there is an additional term $T^0$ with its 
time-dependent coefficient $d(t)$ at the background level. $T^0$ is the 0-index 
component of the contracted torsion tensor $T^{\mu}$, while the full torsion 
tensor is ${T}^{\lambda}{}_{\mu \nu}=h_{a}{}^{\lambda}(\partial_{\mu} h^{a}{}_{ 
\nu}-\partial_{\nu} h^{a}{}_{\mu}+{\omega}^{a}{}_{b \mu} 
h^{b}{}_{\nu}-{\omega}^{a}{}_{b \nu} h^{b}{}_{ \mu})$, with  
$h^{a}{}_{\mu}$ the tetrad field and ${\omega}^{a}{}_{b \mu}$ the spin 
connection which represents inertial effects. Finally, in the above expression 
$S^{(2)}$ contains all operators from the perturbation parts.

From Eq.~\eqref{eq:torEFTact} one can derive  the Friedmann equations 
\begin{align}
\label{eq:f11}
H^2 &= \frac{1}{3 M_{P}^2} \big( \rho_m +\rho_{DE}^{\text{eff}} \big) ~, \\
\dot{H} &= -\frac{1}{2 M_{P}^2} \big( \rho_m +\rho_{DE}^{\text{eff}} +p_m 
+p_{DE}^{\text{eff}} \big) ~,  
\end{align}
where  $\rho_{DE}^{\text{eff}}$ and $p_{DE}^{\text{eff}}$ are the effective 
dark energy density and pressure defined through the operators as 
\citep{Li:2018ixg}
\begin{align}
\label{eq:rhopDE}
\rho_{DE}^{\text{eff}} &= b+\Lambda -3 M_P^2 \Big[ 
H\dot{\Psi}+\frac{dH}{2}+H^2(\Psi-1) \Big] ~, \\
p_{DE}^{\text{eff}} &= b -\Lambda +M_P^2 \Big[ \ddot{\Psi} +2H\dot{\Psi} 
+\frac{\dot{d}}{2} +(H^2 +2\dot{H})(\Psi -1) \Big] ~. 
\label{eq:pppDE}
\end{align}

\subsection{Application of EFT to $f(T)$ Cosmology}

Let us now apply the above EFT approach to torsional gravity in the specific 
case of $f(T)$ cosmology. The general action of 
$f(T)$ gravity is
\begin{align}
S=\int d^{4} x \; e \frac{M_{P}^{2}}{2} [T+f(T)],
\end{align}
where the torsion scalar  is $T=-6H^2$, and $e$ is the determinant of the 
tetrad field $h^{a}{}_{\mu}$, related to the metric through $g_{\mu\nu}=\eta_{ab}\, 
h^{a}{}_{\mu} \, h^{b}{}_{\nu} $ with $\eta_{AB}={\rm diag} 
(1,-1,-1,-1)$). 
Substituting the FLRW tetrad $h^{a}{}_{\mu}={\rm
diag}(1,a,a,a)$, into the field equations we extract  the two modified 
Friedmann equations 
\begin{align}
\label{eq:Frift}
H^{2} &= \frac{8 \pi G}{3} \rho_{m}-\frac{f(T)}{6}+\frac{T f_{T}}{3},  
\\
\dot{H} &= -\frac{4 \pi G\left(\rho_{m}+p_{m}\right)}{1+f_{T}+2 T f_{T T}},
\end{align}
  where   $f_{T}\equiv\partial f/\partial T$ and $f_{TT} \equiv \partial^{2} 
f/\partial T^{2}$.
Comparing them with the standard Friedmann equations we can obtain the 
effective energy density and pressure of dark energy
\begin{align}
\label{eqaux1}
\rho_{f(T)} &=\frac{M_{P}^{2}}{2}\left[2 T f_{T}-f(T)\right],   \\
p_{f(T)} &=\frac{M_{P}^{2}}{2}\left[\frac{f(T)-Tf_{T}+2 T^2 f_{T T}}{1+f_{T}+2 
T 
f_{T T}}\right].  
\label{eqaux2}
\end{align}
Hence, since the above operators in effective 
field theory of $f(T)$ gravity at the background level can be expressed as 
\citep{Li:2018ixg}
\begin{align}
\Psi(t) &=-(1+f_{T}) \  ; \ \
\Lambda(t) =\frac{M_P^2}{2}(T f_T-f), \nonumber \\
d(t) &=2 \dot{f}_{T} \ \  ; \ \ b(t) =0,
\end{align}
expressions Eq.~\eqref{eq:rhopDE} and Eq.~\eqref{eq:pppDE}
coincide with Eq.~\eqref{eqaux1} and Eq.~\eqref{eqaux2}.

As it known, in FLRW geometry, the choice $f(T)=C 
\sqrt{T/T_0}-2\Lambda_{c}$, with $C$  a constant, corresponds to    
$\Lambda$CDM cosmology \citep{Yan:2019gbw}. In this expression $T_0$ 
is the value of $T$ at present, whose existence facilitates the selection of the
units of the various coefficients, and ensures that special terms remain 
consistent under the selection of different metric signatures. In this case, 
the term $\sqrt{T/T_0}$ does not contribute to the effective energy 
density, due  to the cancellation of $\sqrt{T/T_0}$ term in the Friedmann 
equation. Thus, at the background level, the above model is equivalent to 
the universe with a single contribution of a cosmological constant 
$\Lambda_{c}$.

\section{Model-independent reconstructions}
\label{sec:reconstruction}
%

In this section we will apply the method of Gaussian processes in order to show 
how we can reconstruct in a model-independent way an $f(T)$ form in agreement 
with observational data.

\subsection{Gaussian Processes}

The Gaussian processes regression has been widely used to reconstruct 
non-linear functions, and in particular to  obtain the unknown function 
directly from   observational data. The Gaussian processes are a stochastic 
procedure that allows one to acquire a collection of random variables, which 
are subject to a Gaussian distribution \citep{Seikel:2012uu}. The correlation of 
the obtained joint normal distribution function is described by a covariance 
matrix function with specific hyperparameters totally determined 
by data points. Hence, Gaussian processes form a model-independent function 
reconstruction method without any special physical assumption and 
parameterization. Therefore, they are widely used in 
cosmological researches to reconstruct physical parameters from observational 
data sets \citep{Seikel:2013fda, 
Cai:2015zoa, Wang:2017jdm, Elizalde:2018dvw, Aljaf:2020eqh, Holsclaw:2010sk, 
Benisty:2020kdt, Bernardo:2021qhu, Jesus:2021bxq, Rodrigues:2021wyk, 
LeviSaid:2021yat, Cai:2015pia, Mukherjee:2020vkx, vonMarttens:2020apn,Benisty:2022psx}.

In this work we apply GAPP (Gaussian Processes in Python) to reconstruct 
$H(z)$ and their derivatives through observational data points. We will choose 
the exponential kernel form as covariance function, namely
\begin{equation}
 k\left(x, x^{\prime}\right)=\sigma_{f}^{2} 
e^{-\frac{\left(x-x^{\prime}\right)^{2}}{2 l^{2}}},
\end{equation} 
where the $\sigma_{f}$ and $l$ are the hyperparameters.

\subsection{Observational Hubble Data}

We are interested in combining the observational Hubble data (OHD) and the 
latest $H_0$ local measurement to reconstruct the evolutionary history of the
Hubble parameter $H(z)$. The $H(z)$ data are obtained mainly by two methods: 
cosmic chronometer (CC) and radial baryon acoustic oscillations (BAO) 
observations. The cosmic chronometers   provide information of $H(z)$ 
from the age evolution of passively evolving galaxies in a model-independent 
way \citep{Jimenez:2001gg}, while radial BAO measure the clustering of galaxies 
with the BAO peak position as a standard ruler, which depends on the sound 
horizon.

The OHD list has been collected and provided by \citep{Farooq:2016zwm, 
Zhang:2016tto, Yu:2017iju, Magana:2017nfs}. We use the data listed in 
\citep{Li:2019nux}, which includes 31 data points of CC and 23 data points of 
radial BAO. For the value of $H_0$ we use the latest  observation $73.04 \pm 
1.04 \ \mathrm{km\ s^{-1} Mpc^{-1}}$ by SH0ES \citep{Riess:2021jrx}.

\begin{figure}[ht!]
\centering
\includegraphics[width=3.3in]{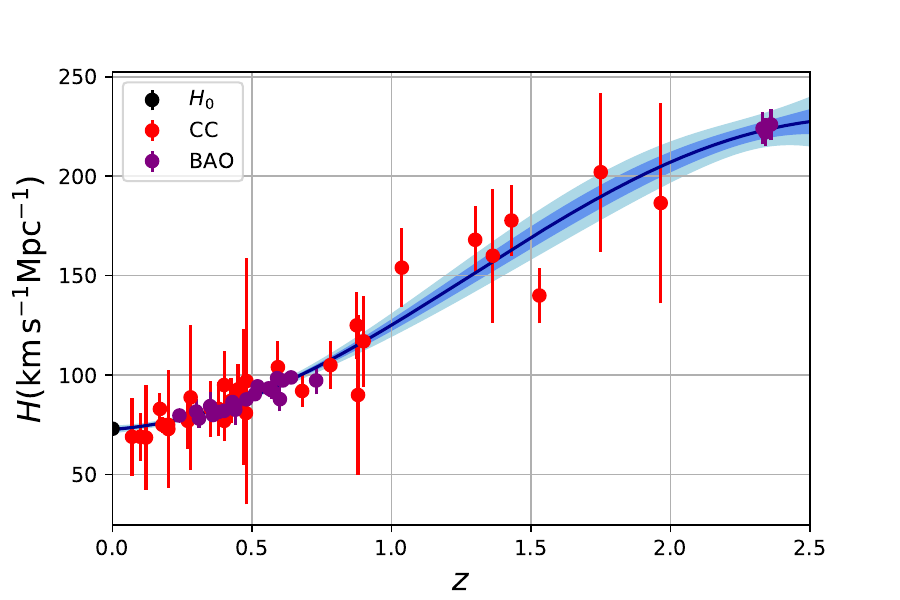} 
\includegraphics[width=3.3in]{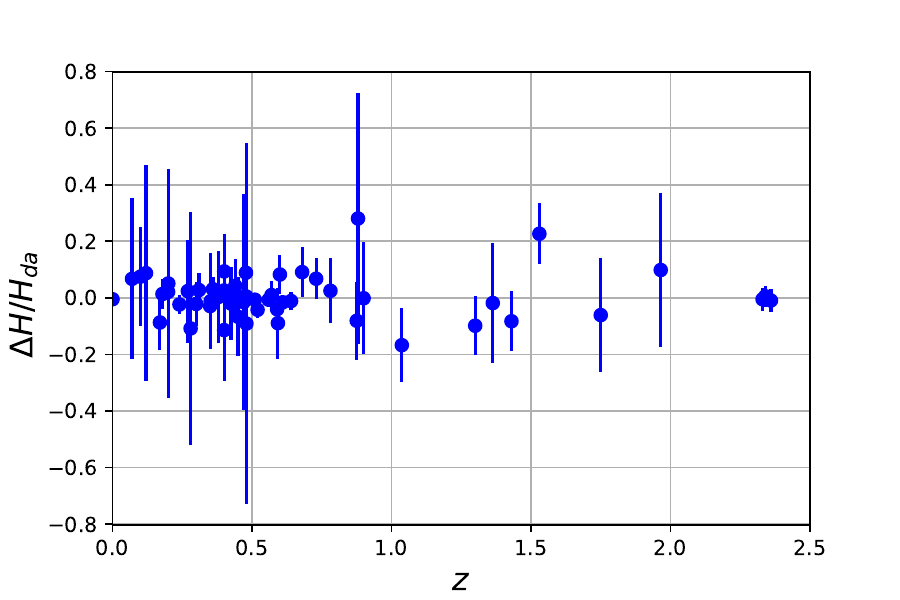} 
\caption{{\it{Upper graph: The reconstructed behavior of $H(z)$, arising from the 
31 CC data 
points, the  23 BAO data points of radial  method, and with $H_0=73.04 \pm 
1.04 km\ s^{-1} Mpc^{-1}$. The dark curve denotes the mean value, while the 
blue and light blur shaded areas mark the allowed regions at $1\sigma$ and 
$2\sigma$ confidence level respectively. {Lower graph: The corresponding residuals quantified 
by the quantity  $\Delta H/{H_{da}}$, where $\Delta 
H=H_{da}-H_{re}$ represents the Hubble parameter difference between the observational value $ H_{da}$ 
and the reconstructed value $H_{re}$.}}}
\label{fig:Hz}}
\end{figure}

With these 55 data points and their error bars, we can reconstruct the redshift 
evolution of $H(z)$ and its derivative. The reconstructed $H(z)$ {and its residuals relative to the data} are shown in 
Fig.~\ref{fig:Hz} with the original observational data points. The dark blue 
curve is the mean value, and the regions of $1\sigma$ and $2\sigma$ confidence 
level are marked in blue and light blue respectively.

{The chi-square values of these 55 points are given by}
\begin{align}
    \chi_{H(z)}^{2}=\sum_{i=1}^{55} \frac{\left(H_{\mathrm{da}, i}-H_{\mathrm{re}, i}\right)^{2}}{\sigma_{H_i}^{2}}=23.885 ~.
\end{align}
{Additionally, we can apply the $\mathcal{R}^{2}$-test to quantify the fitting efficiency between the reconstructed results and data. The coefficient of determination or $\mathcal{R}^{2}$ is defined as \citep{draper1998applied,Capozziello:2017uam}: }
\begin{align}
\mathcal{R}^{2} \equiv 1 -\frac{\sum_{i=1}^{n}\left(H_{\mathrm{da}, i}-H_{\mathrm{re}, i}\right)^{2}}{\sum_{i=1}^{n}\left(H_{\mathrm{da}, i}-\bar{H}\right)^{2}} = 0.9814 ~,
\end{align}
where
\begin{align}
\bar{H} =\frac{1}{n} \sum_{i=1}^{n} H_{\mathrm{da}, i} ~.
\end{align}
{ When the value of $\mathcal{R}^{2}$ is closer to 1, the degree of fitting is better. Therefore, the $H(z)$ obtained through Gaussian processes can conform well  to OHD data.}

\subsection{Reconstruction in the Case of $f(T)$ Cosmology}

In the previous subsection we reconstructed in a model-independent way the 
Hubble function. Hence, we can now use the obtained form in order to 
reconstruct the  $f(T)$ function itself that is responsible for producing it.
Since the torsion scalar in FLRW geometry is only a function of the Hubble 
function and not of its derivative, the whole procedure is significantly easier 
than in other modified gravity theories, such as $f(R)$ gravity. 

The modified 
Friedmann equation  \eqref{eq:Frift} provides the relation between the
$f(T)$ function and $H(z)$. For small $\Delta z$  we can use the approximation  
\begin{align}
    f_{T} \equiv \frac{d f(T)}{d T}=\frac{d f / d z}{d T / d 
z}=\frac{f^{\prime}}{T^{\prime}}, \end{align}
with
\begin{align}
f^{\prime}(z) \approx \frac{f(z+\Delta z)-f(z)}{\Delta z},
\end{align}
and thus $f_{T}$ can be represented by $f(z)$ and $H(z)$. Furthermore, we can 
extract the recursive relation between the consecutive redshifts ($z_{i}$ and 
$z_{i+1}$), namely writing $f(z_{i+1})$ as a function of $f\left(z_{i}\right)$, 
$H(z_i)$ and $H^{\prime}\left(z_{i}\right)$ as
\citep{Cai:2019bdh,Briffa:2020qli,Ren:2021tfi} 
\begin{eqnarray}
	&&
	\!\!\!\!\!\!\!\!\!\!\!\!\!\!
	f\left(z_{i+1}\right)  =f\left(z_{i}\right) +6\left(z_{i+1}-z_{i}\right) 
\frac{H^{\prime}\left(z_{i}\right)}{H\left(z_{i}\right)}  \nonumber\\
	&&\ \ \ \ \ \ \  \cdot \left[H^{2}\left(z_{i}\right)-H_{0}^{2} \Omega_{m 
0}\left(1+z_{i}\right)^{3}+\frac{f\left(z_{i}\right)}{6}\right] .
\label{eq:frecon}
\end{eqnarray}
From this expression we can obtain the value of $f$ at the redshift $z_{i+1}$, 
as long as we know the parameters  at the redshift $z_i$. Finally, from the 
relation between $T$ and $H$, and the evolution of $H(z)$, we can directly 
extract the expression of $f$ as a function of $T$.

\section{Results and features of the reconstructed forms}
\label{sec:features}

In this section we  apply the above procedure and we extract the 
specific $f(T)$ form, investigating its features.

\subsection{Reconstruction Considering Special Function Terms}

\begin{figure}[ht!]
\centering
\includegraphics[width=3.3in]{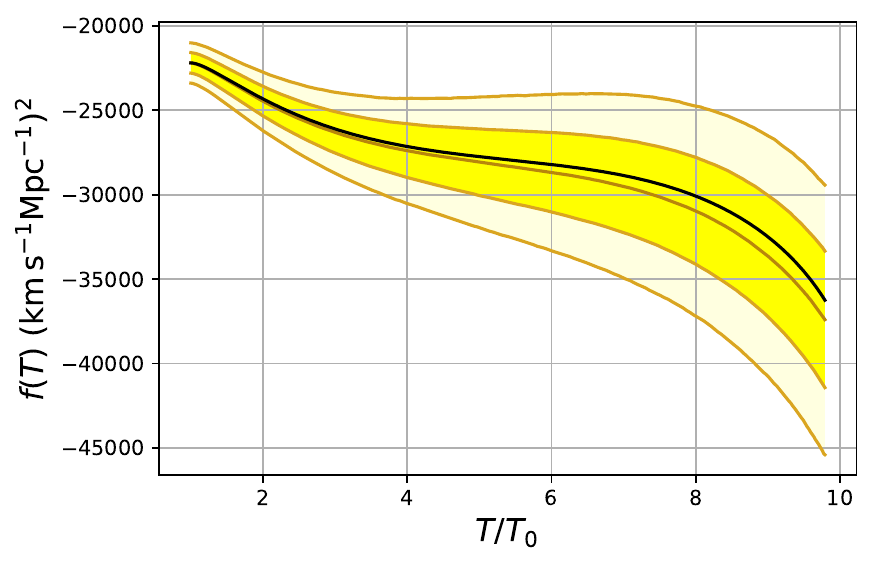}
\caption{\it{The reconstructed $f(T)$ function from observational Hubble data 
through Gaussian processes. The yellow 
and light yellow areas mark the regions of $1\sigma$ and $2\sigma$ confidence 
level respectively, while the black curve arises using the best-fit curve of 
$f(T)$. Both $T$ and $f(T)$ are measured in units of  $(\mathrm{km} \, 
\mathrm{s}^{-1} \mathrm{Mpc}^{-1})^2$ and $T_0=-6H_0^2$.}}
\label{fig:ffromH}
\end{figure}

With the model-independent reconstruction method introduced in 
Section \ref{sec:reconstruction}, we can reconstruct $f(T)$   by fitting $H(z)$ 
and $H'(z)$. As we described, the observational Hubble data directly determine 
$H(z)$, and thus by applying the Gaussian processes we reconstruct $H'(z)$ 
too. Hence, we can finally use Eq.~\eqref{eq:frecon} and  reconstruct the 
corresponding  $f(T)$ form with $\Omega_{m0}=0.3$ presented in   Fig.~\ref{fig:ffromH}, where $T_0$ is 
the value of $T$ at present. When $f(T)=-2\Lambda_c$ we recover  $\Lambda$CDM 
cosmology. Note that the units of both $T$ and $f(T)$ are $(\mathrm{km} \, 
\mathrm{s}^{-1} \mathrm{Mpc}^{-1})^2$.

\begin{figure}[ht!]
\centering
\includegraphics[width=3.3in]{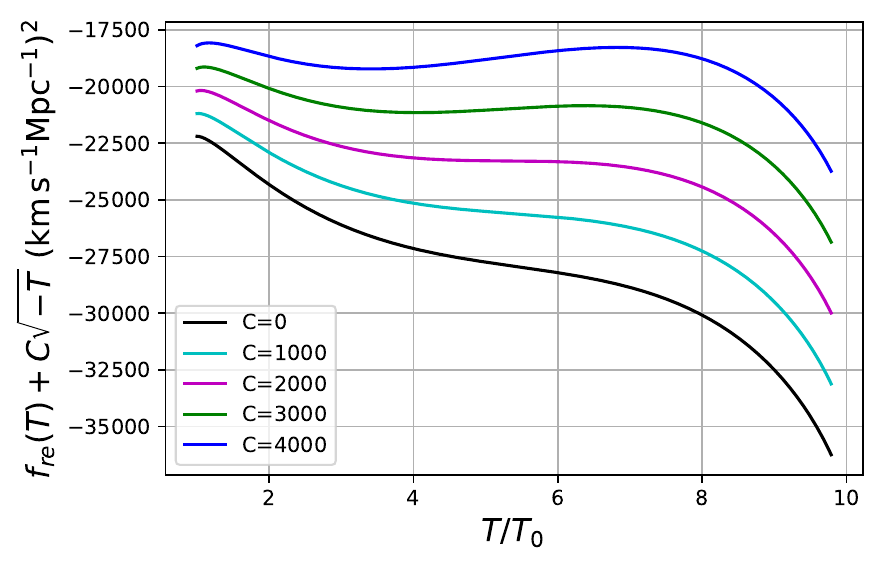}
\caption{\it{The reconstructed general $f(T)$   function $f_{ge}(T)=f_{re}(T)+C 
\sqrt{T/T_0}$, with  $f_{re}(T)$ the reconstructed mean curve of  
Fig.~\ref{fig:ffromH}, with $T_0=-6H_0^2$ and different choices of the 
parameter $C$.  Both $T$ and $f(T)$, as well as $C$, are measured in 
units of  $(\mathrm{km} \, 
\mathrm{s}^{-1} \mathrm{Mpc}^{-1})^2$.}}
\label{fig:sqrtC}
\end{figure}

Nevertheless, let us note here  that the reconstructed $f(T)$ form  in  
Fig.~\ref{fig:ffromH} is not the only function that can provide the 
corresponding evolution of $H(z)$ of Fig.~\ref{fig:Hz}, since as we mentioned 
above   the addition of the special term $\sqrt{T/T_0}$ will not affect the 
evolution at the background level. Thus,  for any $f_{re}(T)$ form that is 
reconstructed using the Gaussian processes, there is an extra possibility, 
namely a more general  $f_{ge}(T)=f_{re}(T)+C 
\sqrt{T/T_0}$, where $C$ is an arbitrary coefficient, that will produce  the 
same evolution of $H(z)$. This behavior is depicted in    Fig.~\ref{fig:sqrtC}, 
where the different $f(T)$  
forms correspond to the same background 
evolution of the universe.
 \begin{figure}[ht!]
\centering
\includegraphics[width=3.3in]{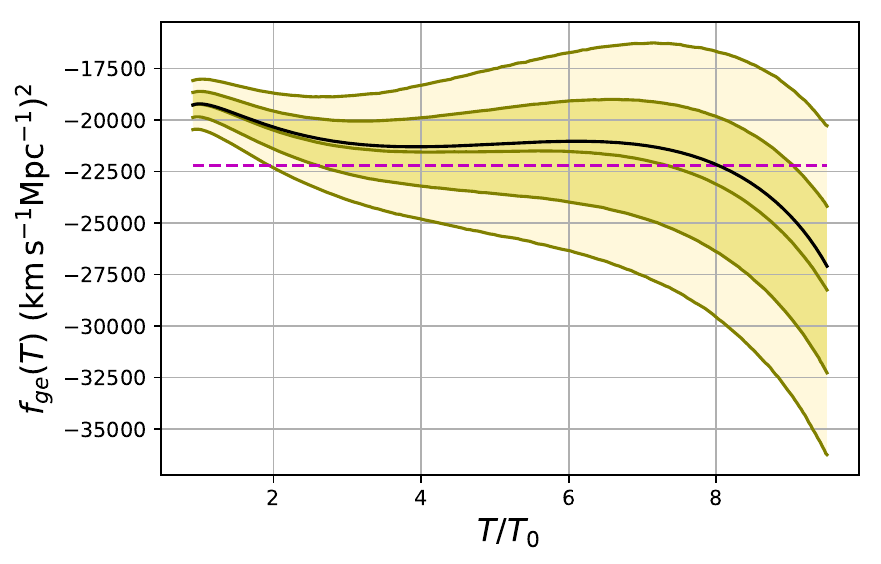}
\caption{\it{The reconstructed general $f(T)$   function $f_{ge}(T)=f_{re}(T)+C 
\sqrt{T/T_0}$, with  $f_{re}(T)$ the reconstructed result of  
Fig.~\ref{fig:ffromH}, in the specific case where $C=2927$ compared to the 
$\Lambda$CDM model (red dotted line). The latter  
  lies within the $1\sigma$ reconstructed region  for a relatively large time 
range.  Both $T$ and $f(T)$, as well as $C$, 
are measured in 
units of  $(\mathrm{km} \, 
\mathrm{s}^{-1} \mathrm{Mpc}^{-1})^2$ and $T_0=-6H_0^2$.}}
\label{fig:sqrtCDM}
\end{figure}

We can now use the above features in order to  extract the reconstructed 
function that is closest to the $\Lambda$CDM cosmology bust still be in 
agreement with the data. As we show in Fig.~\ref{fig:sqrtCDM}, applying the 
curve fitting method  we find that when the coefficient is $C=2927$ the 
reconstructed function is closest to the cosmological constant, or equivalently 
in this case the 
cosmological constant   lies within the $1\sigma$ region of the reconstructed 
$f(T)$ model for a relatively large time range.

\subsection{Analytic Fittings of the Reconstructed Function}

In order to describe the reconstructed $f(T)$ function and its corresponding 
cosmological parameters in a more accurate way, we use polynomial fittings to 
extract analytic forms.
The function fitting is based on the 251 reconstructed $f(T)$ points, 
while the original data source contains $H(z)$. Through Gaussian processes 
and approximate reconstruction, we can transform the information of $H(z)$ into the evolution of $f(T)$, and thus we can find the appropriate analytical expression of $f(T)$. In order to quantify the efficiency of the fitting of the various models, we need to implement particular information criteria.

 The Bayesian Information Criterion (BIC) is the Criterion to select the model with the best fitting behavior among many models, and the model with the lower BIC  is statistically favored \citep{Liddle:2007fy,Anagnostopoulos:2019miu}. BIC is defined as
\begin{align}
    \mathrm{BIC}=-2 \ln \left(\mathcal{L}_{\max }\right)+k \ln \left(N_{\text {tot }}\right) ,
\end{align}
where $\mathcal{L}_{\max }$ is the maximized value of the likelihood function
of the specific model, $N_{\text {tot }}$ is the number of data points, and  $k$
 the number of the model parameters. 
In  Tab.\ref{tab:bic} we summarize the BIC values of the standard model  $f(T) = - 2\Lambda  +a_0 \sqrt{T/T_0} + a_1 T$,  the best fit quadratic polynomial $f(T)$ model $f(T) = - 2\Lambda  +a_0 \sqrt{T/T_0} + a_1 T +a_2 T^2$, the best fit cubic polynomial $f(T)$ model $f(T) = - 2\Lambda  +a_0 \sqrt{T/T_0} + a_1 T +a_2 T^2 + a_3 T^3$, as well as  the best fit quartic polynomial $f(T)$ model $f(T) = - 2\Lambda  +a_0 \sqrt{T/T_0} + a_1 T +a_2 T^2 + a_3 T^3+ a_4 T^4$, compared to the reconstructed result. Since the parameter $a_0$ does not affect the evolution of $H(z)$, and $a_1$ can be eliminated by parameter rescaling, these two parameters are not effective degrees of freedom. Standard model and quadratic polynomial model are not efficient to quantify  the reconstructed $f(T)$ functions. On the other hand,  quartic polynomial describes the reconstructed functions well, nevertheless the extra parameters cause its BIC to be less good than cubic polynomial. As we deduce from BIC, the cubic polynomial $f(T)$  is the best fitting model.

\begin{table} \label{tab:bic}
\begin{center}
\begin{tabular}{c|c|c}
\hline\hline
	Model	         & $BIC$ & $\Delta BIC$ \\   \hline
$\Lambda$CDM & 42.54 & 14.49 \\	    
 quadratic polynomial $f(T)$ model & 37.28 & 9.23  \\
cubic polynomial $f(T)$ model & 28.05 & 0  \\
quartic polynomial $f(T)$ model & 33.57 & 5.52  \\ 
\hline\hline
\end{tabular}
\end{center}
\caption{The BIC values for the examined $f(T)$ models, alongside   the corresponding
differences $\Delta BIC = BIC-BIC_{min}$.}
\end{table}

\begin{figure}[ht]
    \centering
    \includegraphics[width=3.3in]{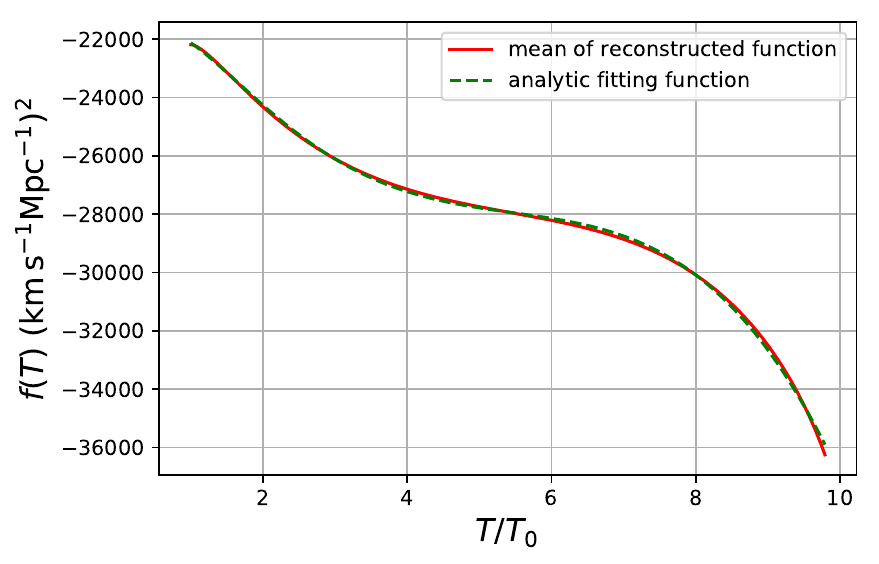}
    \includegraphics[width=3.3in]{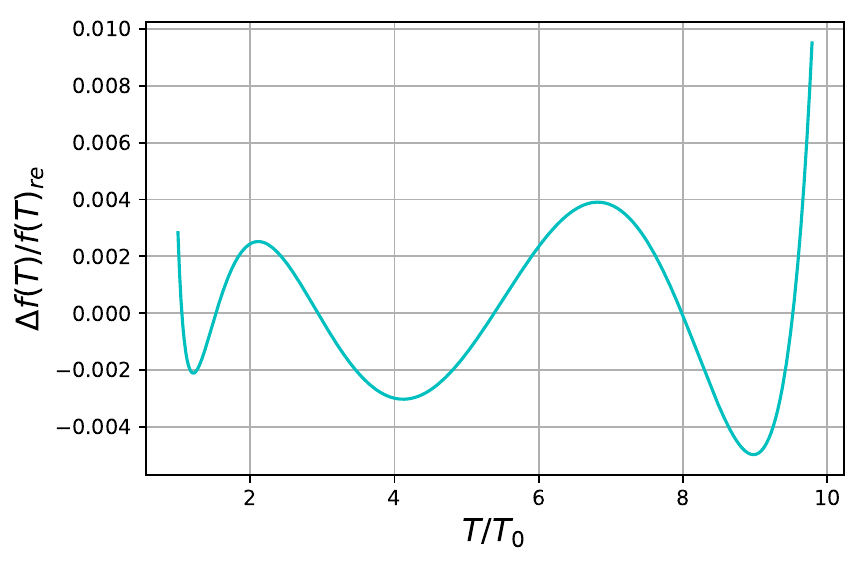}
    \caption{\it{Upper graph: The best fit analytical model of $f(T)$ gravity (dashed 
purple curve), namely  $f(T) = - 2\Lambda  +a_0 \sqrt{T/T_0} + a_1 T +a_2 T^2 + a_3 T^3$, with $a_3=-93.50$, 
$a_2=1750$, $a_1=-15650$, $a_0=21570$ and $\Lambda=14850$ in 
units of  $(\mathrm{km} \, 
\mathrm{s}^{-1} \mathrm{Mpc}^{-1})^2$, and the  mean curve 
of the  model-independent reconstructed $f(T)$ from OHD (orange curve). 
{Lower graph: The corresponding differences quantified 
by the quantity  $\Delta f(T)/{f(T)_{re}}$, where $\Delta 
H=f(T)_{re}-f(T)_{th}$ represents the  Hubble parameter difference between the 
reconstructed value $f(T)_{re}$ and the analytical value $f(T)_{th}$.} }}
    \label{fig:fTvsT}
\end{figure}

In summary, we chose cubic polynomials, including the
mentioned square-root term $\sqrt{T/T_0}$, to describe the reconstructed $f(T)$ function analytically, namely:
\begin{align}
f(T) = - 2\Lambda  +a_0 \sqrt{T/T_0} + a_1 T +a_2 T^2 + a_3 T^3,
\label{eq:analytic}
\end{align}
and the fitting result is depicted in Fig.~\ref{fig:fTvsT}. The analytic 
function (dashed green curve) can describe the model-independent reconstructed 
$f(T)$ form (red curve) very efficiently by choosing  $a_3=-93.50$, 
$a_2=1750$, $a_1=-15650$, $a_0=21570$ and $\Lambda=14850$ in 
units of  $(\mathrm{km} \, 
\mathrm{s}^{-1} \mathrm{Mpc}^{-1})^2$ .
\begin{figure}
\centering
\includegraphics[width=3.3in]{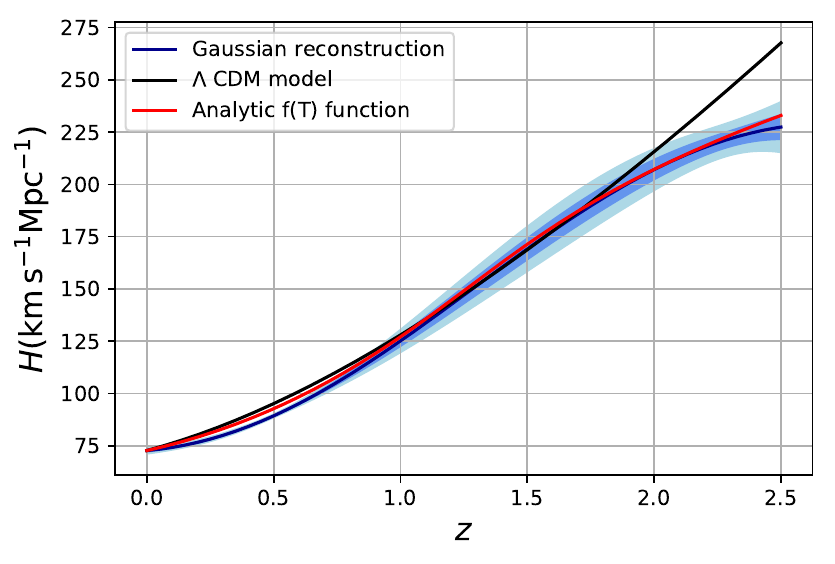}    
\includegraphics[width=3.3in]{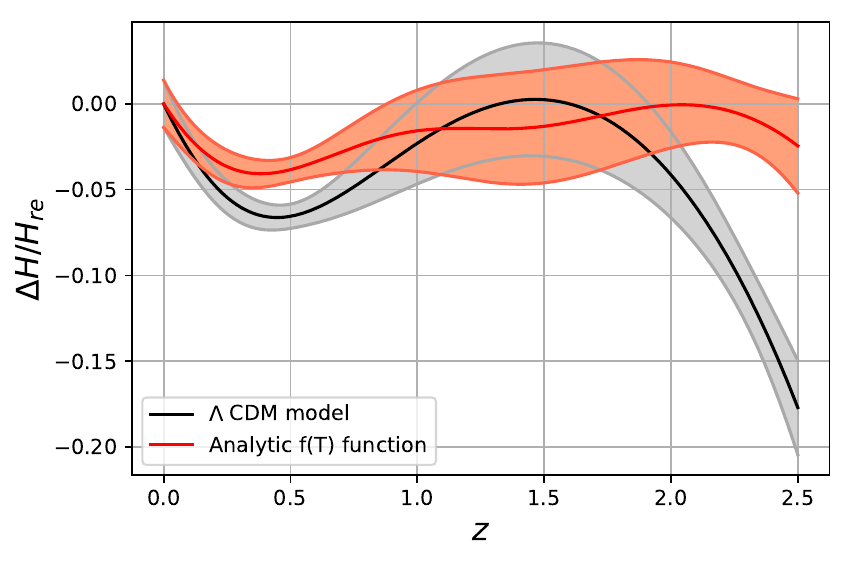} 
\caption{{\it{Upper graph: The   $H(z)$ evolution    obtained 
from the Gaussian process reconstruction  (blue curves and shadows),  from the 
analytical expression  \eqref{eq:analytic}  (red curve), and from  $\Lambda$CDM 
scenario (black curve). Lower graph: The corresponding differences quantified 
by the quantity  $\Delta H/{H_{re}}$, where $\Delta 
H=H_{re}-H_{th}$ represents the    Hubble parameter difference between the 
reconstructed value $ H_{re}$ and the theoretical value $ H_{th}$.}}
\label{fig:hcompare}}
\end{figure}

As a cross-test of the above procedure we proceed as follows. Since we have 
obtained the  fittings of the analytic expression of the $f(T)$ function, we can 
insert Eq.~\eqref{eq:analytic} into the modified Friedmann equation 
\eqref{eq:Frift} and extract the solution for $H(z)$. In the upper graph of 
Fig.~\ref{fig:hcompare} we present the results, namely the $H(z)$ obtained 
from the Gaussian process reconstruction and the $H(z)$ obtained  from the 
analytical expression  \eqref{eq:analytic}, and for completeness we add also 
the   $H(z)$  corresponding to $\Lambda$CDM cosmology.
Additionally, in order to examine the corresponding differences  we introduce  
the quantity  $\Delta H/{H_{re}}$, where $\Delta 
H=H_{re}-H_{th}$ represents the    Hubble parameter difference between the 
reconstructed value $ H_{re}$ and the theoretical value $ H_{th}$. 
In 
the the lower graph of Fig.~\ref{fig:hcompare}  we present its behavior, and as 
we can see, compared to  $\Lambda$CDM model  the analytical form  
\eqref{eq:analytic}  produces a result closer to that  of data 
reconstruction.

\section{Conclusion and Discussions}
\label{sec:conclusion}

In this work we considered the effective field theory of $f(T)$ gravity as a 
framework to study the background evolution of the universe. We used 
the  latest observational Hubble data and the $H_0$ measurement,  and we 
reconstructed the Hubble function $H(z)$ by applying    Gaussian processes. 
Then we used the obtained form in order to reconstruct the $f(T)$ function in a 
model-independent way. Since the special term 
 $\sqrt{T/T_0}$ does not affect the evolution of the 
cosmological background,  the family of all functions  
$f_{ge}(T)=f_{re}(T)+C \sqrt{T/T_0}$  produces the same background evolution 
with the reconstructed $f_{re}(T)$ form.

Having the $f(T)$ form obtained from data reconstruction, we performed a 
  fitting by using polynomial functions containing additionally the square root 
term. We found that the reconstructed $f(T)$ expression presented in 
Eq.~\eqref{eq:analytic} may describe observations more efficiently  than the 
$\Lambda$CDM scenario. This is the main result of the present work.  

Furthermore, we expressed the Friedmann equations under $f(T)$ gravity as the recursive expression of $f(z)$, applying   different approximation methods.
This can be well combined with the $H(z)$ obtained through the Gaussian processes, in order to obtain
the reconstruction of $f(T)$ function. We mention here that there are alternative  procedures to reconstruct the $f(T)$ function.
Specifically, the redshift $z$ can be expressed in terms of Hubble, deceleration, jerk and snap parameters,
through the back scattering approach. Combined with a general function that sets $f$ to $z$, the
reconstruction of $f(T)$ can be achieved \citep{Capozziello:2017uam}. 
This reconstruction method is suitable when we have  information at $z=0$.  Moreover, 
it is   an effective method to use other polynomials to approximate $f(z)$
and perform parameter fittings \citep{Capozziello:2019cav}. 
Using suitable polynomials can   achieve the same effect as the $f(z)$ recursive expressions of the present work. 

Finally, we mention that the above analysis was based on  data by 
HST and SH0ES teams, however since  the tension between 
different datasets increases \citep{Riess:2021jrx}, it would be interesting to  extend it using the 
Pantheon$+$ Type Ia supernovae (SNIa) data  
\citep{Brout:2022vxf,Scolnic:2021amr,Brownsberger:2021uue}.  Additionally, as mentioned above,   the 
 $a_0\sqrt{T/T_0}$ term  does not affect the  background cosmological evolution, 
however  at the perturbative level, and in particular at the evolution equation 
of the matter overdensity, this term will have an effect since different $a_0$ 
values will  lead to different gravitational constant 
$G_{\text{eff}}={G_N}/{(1+f_T)}$ \citep{Anagnostopoulos:2019miu}. 
Hence, although    this term does not 
affect the cosmological background, and hence the Gaussian processes procedure, 
it does affect the  evolution of perturbations. In summary, it would be both 
interesting and necessary to perform the Gaussian process analysis using 
additionally SNIa as well as growth data, in order to obtain more accurate 
results and break the degeneracies between different $f(T)$ forms. Since these 
investigations lie beyond the scope of the present work, are left for future 
projects.

\section*{Acknowledgments}

We are grateful to Amara Ilyas, Geyu Mo , Wentao Luo, Dongdong Zhang and Sunny Vagnozzi for helpful 
discussions. This work is supported in part by the National Key R\&D Program of 
China (2021YFC2203100), by the NSFC (11961131007, 11653002), by the Fundamental 
Research Funds for Central Universities, by the CSC Innovation Talent Funds, by 
the CAS project for young scientists in basic research (YSBR-006), by the USTC 
Fellowship for International Cooperation, and by the USTC Research Funds of the 
Double First-Class Initiative. SFY is partially supported by the Disposizione del Presidente INFN n.21786 ``{\it Quantum Fields for Gravity, Cosmology and Black Holes}''.
ENS acknowledges participation in the COST Association Action CA18108 ``{\it 
Quantum Gravity Phenomenology in the Multimessenger Approach (QG-MM)}''.
All numerics were operated on the computer clusters {\it LINDA} \& {\it JUDY} in 
the particle cosmology group at USTC.

\bibliography{main}{}
\bibliographystyle{aasjournal}

\end{document}